\documentclass[12pt, twocolumn, english]{iopart}
\expandafter\let\csname equation*\endcsname\relax
\expandafter\let\csname endequation*\endcsname\relax
\usepackage{amsmath}
\usepackage[english]{babel}
\usepackage{bm}
\usepackage{bbm}
\usepackage{iopams}
\usepackage{graphicx}

\usepackage{color}
\newcommand{\blue}{\color{blue}}

\usepackage[%
  breaklinks=true,
  colorlinks=true,
  linkcolor=blue,
  urlcolor=blue,
  citecolor=blue
]{hyperref}
 \usepackage[numbers,sort&compress]{natbib}
 \usepackage{hypernat}
 
\begin{document}
\title[]{Generalisation of Gilbert damping and magnetic inertia parameter as a series of higher-order relativistic terms}

\author{Ritwik Mondal\footnote{Present address: Department of Physics, University of Konstanz, D -78457 Konstanz, Germany}, Marco Berritta and  Peter M. Oppeneer}

\address{Department of Physics and Astronomy, Uppsala University, P.\,O.\ Box 516, SE-751\,20 Uppsala, Sweden}
\ead{ritwik.mondal@physics.uu.se}

\begin{abstract}
  The phenomenological Landau-Lifshitz-Gilbert (LLG) equation of motion remains as the cornerstone of contemporary magnetisation dynamics studies, wherein the Gilbert damping parameter has been attributed to first-order relativistic effects. To include  magnetic inertial effects the  LLG equation has previously been extended with a supplemental inertia term and the arising inertial dynamics has been related to second-order relativistic effects. 
 Here we start from the relativistic Dirac equation and, performing a Foldy-Wouthuysen transformation, derive a generalised Pauli spin Hamiltonian that contains relativistic correction terms to any higher order. Using the Heisenberg equation of spin motion we derive general relativistic expressions for the tensorial Gilbert damping and magnetic inertia parameters, and show that these tensors can be expressed as series of higher-order relativistic correction terms. We further show  that, in the case of a harmonic external driving field, these series can be summed and we provide closed analytical expressions for the Gilbert and inertial parameters that are functions of the frequency of the driving field.   \\
 \end{abstract}

\section{Introduction}
Spin dynamics in magnetic systems has often been described by the phenomenological Landau-Lifshitz (LL) equation of motion of the following form \cite{landau35} 
\begin{equation}
\frac{\partial \bm{M}}{\partial t} =- \gamma  \bm{M} \times \bm{H}_{\rm eff}   -\lambda \bm{M} \times [\bm{M} \times \bm{H}_{\rm eff}],
\label{LL}
\end{equation}
where $\gamma$ is the gyromagnetic ratio, $\bm{H}_{\rm eff}$ is the effective magnetic field, and $\lambda$ is an isotropic damping parameter. The first term 
describes the precession of the local, classical magnetisation vector $\bm{M} (\bm{r}, t)$ around the effective field $\bm{H}_{\rm eff}$.
The second term describes the magnetisation relaxation such that the magnetisation vector relaxes to the direction of the effective field until finally it is aligned with the effective field. To include large damping, the relaxation term in the LL equation was reformulated by Gilbert \cite{gilbert04,gilbert56} to give the Landau-Lifshitz-Gilbert (LLG) equation,
\begin{equation}
\frac{\partial \bm{M}}{\partial t} =- \gamma  \bm{M} \times \bm{H}_{\rm eff}   + \alpha  \,  \bm{M} \times \frac{\partial \bm{M}}{\partial t },
\label{LLG}
\end{equation}
where $\alpha$ is the Gilbert damping constant. Note that both damping parameters $\alpha$ and $\lambda$ are here scalars, which corresponds to the assumption of an isotropic medium. Both the LL and LLG equations preserve the length of the magnetisation during the dynamics and are mathematically equivalent (see, e.g.\ \cite{lakshmanan11}). 
\begin{figure}[h!]
	\centering
	\includegraphics{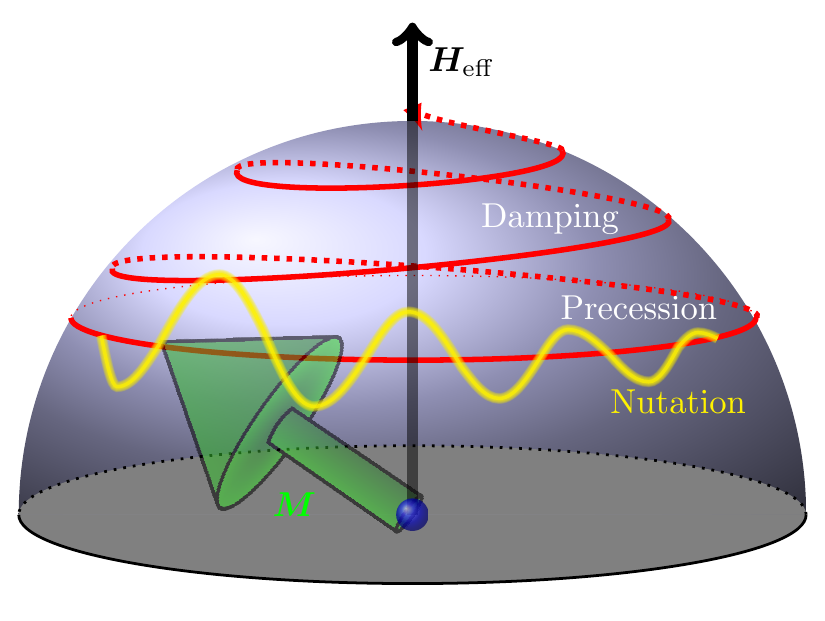}
	\caption{Sketch of extended LLG magnetisation dynamics. The green arrow denotes the classical magnetisation vector which precesses around an effective field. The red solid and dotted lines depict the precession and damping. The yellow path signifies the nutation, or inertial damping, of the magnetisation vector.}
	\label{mag-dyn}
\end{figure}
Recently, there have also been attempts to investigate the magnetic inertial dynamics which is essentially an extension to the LLG equation with an additional term \cite{Wegrowe2012,Wegrowe2015JAP,Wegrowe2016JPCM}. Phenomenologically this additional term of magnetic inertial dynamics, $\bm{M} \times {\mathcal{I}}\,  \partial^2 \bm{M}/ \partial t^2$, can be seen as a torque due to second-order time derivative of the magnetisation \cite{Ciornei2011,Kimel2009,Bhattacharjee2012,Fahnle2011}.  
The essence of the terms in the extended LLG equation is described pictorially in Fig.\ \ref{mag-dyn}. 
Note that in the LLG dynamics the magnetisation is described as a classical vector field and not as a quantum spin vector.

In their original work, Landau and Lifshitz attributed the damping constant $\lambda$ to  relativistic origins \cite{landau35}; later on, it has been more specifically attributed to spin-orbit coupling \cite{kunes02,Pelzl2003,hickey2009,He2013}. 
In the last few decades, several explanations have been proposed towards the origin of damping mechanisms, e.g., the  breathing Fermi surface model \cite{kambersky70,kambersky76}, torque-torque correlation model \cite{kambersky07}, scattering theory formulation \cite{Brataas2008}, effective field theories \cite{Fahnle2011JPCM} etc. 
On the other hand, the origin of magnetic inertia is less discussed in the literature, although it's application to ultrafast spin dynamics and switching could potentially be rich \cite{Kimel2009}. To account for the magnetic inertia, the breathing Fermi surface model has been extended \cite{Fahnle2011,fahnle2013erratum} and the inertia parameter has been associated with the magnetic susceptibility \cite{Thonig2017}.
However, the microscopic origins of both Gilbert damping and magnetic inertia are still under debate and pose a fundamental question that requires to be further investigated. 

In two recent works \cite{Mondal2016,Mondal2017Nutation}, we have shown that both quantities are of relativistic origin. In particular, we derived the Gilbert damping dynamics from the relativistic spin-orbit coupling and showed that the damping parameter is not a scalar quantity but rather a tensor that involves two main contributions: electronic and magnetic ones \cite{Mondal2016}. The electronic contribution is calculated as an electronic states' expectation value of the product of different components of position and momentum operators; however, the magnetic contribution is given by the imaginary part of the susceptibility tensor. 
In an another work, we have derived the magnetic inertial dynamics from a higher-order ($1/c^4$) spin-orbit coupling and showed that the corresponding parameter is also a tensor which depends on the real part of the susceptibility \cite{Mondal2017Nutation}. Both these investigations used a semirelativistic expansion of the Dirac Hamiltonian employing the Foldy-Wouthuysen transformation to obtain an extended Pauli Hamiltonian including the relativistic corrections \cite{Mondal2015a,Mondal2017thesis}. The thus-obtained semirelativistic Hamiltonian was then used to calculate the magnetisation dynamics, especially for the derivation of the LLG equation and magnetic inertial dynamics.         

In this article we use an extended approach towards a derivation of the generalisation of those two (Gilbert damping and magnetic inertia) parameters from the relativistic Dirac Hamiltonian, 
developing a series to fully include the occurring  higher-order relativistic terms.
To this end we start from the Dirac Hamiltonian in the presence of an external electromagnetic field and derive a semirelativistic expansion of it. By doing so, we consider the direct field-spin coupling terms and show that these terms can be written as a series of  higher-order relativistic contributions. Using the latter Hamiltonian, we derive the corresponding spin dynamics. Our results show that the Gilbert damping parameter and inertia parameter can be expressed as a convergent series of higher-order relativistic terms and we derive closed expressions for both quantities. At the lowest order, we find exactly the same tensorial quantities that have been found in earlier works.

\section{Relativistic Hamiltonian Formulation}
To describe a relativistic particle, we start with a Dirac particle \cite{Dirac1928} inside a material, and, in the presence of an external field, for which one can write the Dirac equation as $i\hbar\frac{\partial \psi(\bm{r},t)}{\partial t} = \mathcal{H}\psi(\bm{r},t)$ for a Dirac bi-spinor $\psi$. Adopting furthermore the relativistic density functional theory (DFT) framework we write the corresponding Hamiltonian as \cite{Mondal2015a,Mondal2016,Mondal2017Nutation}
\begin{align}
\label{Dirac_Hamiltonian}
\mathcal{H}&= c\,\underline{\bm{\alpha}}\cdot\left(\bm{p}-e\bm{A}\right)+(\underline{\beta}-\underline{\mathbbm{1}}) mc^{2}+ V \underline{\mathbbm{1}} \nonumber\\
& = \mathcal{O}+(\underline{\beta}-\underline{\mathbbm{1}}) mc^{2}+\mathcal{E} ,
\end{align}
where $V$ is the effective unpolarised Kohn-Sham potential created by the ion-ion, ion-electron and electron-electron interactions.
Generally, to describe magnetic systems, an additional spin-polarised energy (exchange energy) term is required. However, we have treated effects of the exchange field previously, and since it doesn't contribute to the damping terms  we do not consider it explicitly here  (for details of the calculations involving the exchange potential, see Ref.\ \cite{Mondal2015a,Mondal2016}).   
The effect of the external electromagnetic field has been accounted through the vector potential, $\bm{A}(\bm{r},t)$, $c$ defines the speed of light, $m$ is particle's mass and $\underline{\mathbbm{1}}$ is the $4\times 4$ unit matrix.  
 $\underline{\bm{\alpha}}$ and $\underline{\beta}$  are the Dirac matrices which have the form
\[
\underline{\bm{\alpha}}=\left(\begin{array}{cc}
0 & \bm{\sigma}\\
\bm{\sigma} & 0
\end{array}\right),\quad\mbox{}\quad\underline{\beta}=\left(\begin{array}{cc}
\bm{1} & 0\\
0 & -\bm{1}
\end{array}\right)\,,
\]
where $\bm{\sigma} = (\sigma_x,\sigma_y,\sigma_z)$ are the Pauli spin matrix vectors and $\bm{1}$ is $2\times 2$ unit matrix. 
Note that the Dirac matrices form the diagonal and off-diagonal matrix elements of the Hamiltonian in Eq.\ (\ref{Dirac_Hamiltonian}). For example, the off-diagonal elements can be denoted as 
$\mathcal{O}=c\underline{\bm{\alpha}}\cdot\left(\bm{p}-e\bm{A}\right)$, and the diagonal matrix elements can be written as $\mathcal{E}=V \underline{\mathbbm{1}}$.

In the nonrelativistic limit, the Dirac Hamiltonian equals the Pauli Hamiltonian, see e.g.\ \cite{strange98}. In this respect, one has to consider that the Dirac bi-spinor can be written as 
\[\psi(\bm{r},t) = \left(\begin{array}{cc}
\phi(\bm{r},t)\\
\eta(\bm{r},t) 
\end{array} \right), \]
where the upper $\phi$ and lower $\eta$ components have to be considered as ``large'' and ``small'' components, respectively.
This nonrelativistic limit is only valid for the case when the particle's momentum is much smaller than the rest mass energy, otherwise it gives  an unsatisfactory result \cite{Mondal2017thesis}. 
Therefore, the issue of separating the wave functions of particles from those of antiparticles is not clear for any given momentum. This is mainly because the off-diagonal Hamiltonian elements link the particle and antiparticle. 
The Foldy-Wouthuysen (FW) transformation \cite{foldy50} has been a very successful attempt to find a representation where the off-diagonal elements have been reduced in every step of the transformation.
Thereafter, neglecting the higher-order off-diagonal elements, one finds the correct Hamiltonian that describes the particles efficiently.
The FW transformation is an unitary transformation obtained by suitably choosing the FW operator \cite{foldy50}, 
\begin{align}
\label{unitary_operator}
U_{\rm FW} &= -\frac{i}{2mc^2}\underline{\beta}\mathcal{O}.
\end{align}
The minus sign in front of the operator is because of the property that $\underline{\beta}$ and $\mathcal{O}$ anticommute with each other.
With the FW operator, the FW transformation of the wave function adopts the form $\psi^{\prime}(\bm{r},t) = e^{iU_{\rm FW}}\psi(\bm{r},t)$ such that the probability density remains the same, $\vert \psi\vert^2 = \vert\psi^{\prime} \vert^2$. 
In this way, the time-dependent FW transformed Hamiltonian can be expressed as \cite{strange98,Silenko2016PRA,Mondal2017thesis}
\begin{align}
\label{FW_transformation}
\mathcal{H}_{\rm FW}&= e^{iU_{\rm FW}}\left(\mathcal{H}-i\hbar\frac{\partial}{\partial t}\right)e^{-iU_{\rm FW}} + i\hbar \frac{\partial}{\partial t}\,.
\end{align}
According to the Baker-Campbell-Hausdorff formula, the above transformed Hamiltonian can be written as a series of commutators, and the finally transformed Hamiltonian reads
\begin{align}
	\mathcal{H} _{\rm FW} & = \mathcal{H} + i\left[U_{\rm FW},\mathcal{H} - i\hbar\frac{\partial }{\partial t}\right] +\frac{i^2}{2!}\left[U_{\rm FW},\left[U_{\rm FW},\mathcal{H} - i\hbar\frac{\partial }{\partial t}\right]\right]\nonumber\\
	& + \frac{i^3}{3!}\left[ U_{\rm FW},\left[U_{\rm FW},\left[U_{\rm FW},\mathcal{H} - i\hbar\frac{\partial }{\partial t}\right]\right]\right]+ ....\,\,.
	\label{trans-Hamil}
\end{align}
In general, for a time-independent FW transformation, one has to work with $\frac{\partial U_{\rm FW}}{\partial t} = 0$. However, this is only valid if the odd operator does not contain any time dependency. In our case, a time-dependent transformation is needed as the vector potential is notably time-varying. In this regard, we notice that the even operators and the term $i\hbar\,\partial/\partial t$ transform in a similar way. Therefore, we define a term $\mathcal{F}$ such that $\mathcal{F}= \mathcal{E} -i\hbar\, \partial/\partial t $. The main theme of the FW transformation is to make the odd terms smaller in every step of the transformation. After a fourth transformation and neglecting the higher order terms, the Hamiltonian with only the even terms can be shown to have the form as \cite{Mondal2017thesis,schwabl2008advanced,Silenko2016PRA,Silenko2013,Silenko2016}
\begin{align}
\mathcal{H}_{\rm FW} ^{\prime\prime\prime} & = (\underline{\beta} -\underline{\mathbbm{1}})mc^2 + \underline{\beta}\left(\frac{\mathcal{O}^2}{2mc^2}-\frac{\mathcal{O}^4}{8m^3c^6} + \frac{\mathcal{O}^6}{16m^5c^{10}}\right) + \mathcal{E} -  \frac{1}{8m^2c^4}\left[\mathcal{O},\left[\mathcal{O},\mathcal{F}\right]\right]  \nonumber\\
& - \frac{\underline{\beta}}{8m^3c^6}\left[\mathcal{O},\mathcal{F}\right]^2 + \frac{3}{64m^4c^8}\left\{\mathcal{O}^2,\left[\mathcal{O},\left[\mathcal{O},\mathcal{F}\right]\right]\right\}+ \frac{5}{128m^4c^8}\left[\mathcal{O}^2,\left[\mathcal{O}^2,\mathcal{F}\right]\right]\,. 
\label{originalHFW}
\end{align}
Here, for any two operators $A$ and $B$ the commutator is defined as $[A,B]$ and the anticommutator as $\left\{A,B\right\}$. 
As already pointed out, the original FW transformation can only produce correct and expected higher-order terms up to first order i.e., $1/c^4$ \cite{Mondal2017thesis,Silenko2016,Silenko2016PRA}. In fact, in their original work Foldy and Wouthuysen derived only the terms up to $1/c^4$, i.e., only the terms in the first line of Eq.\ (\ref{originalHFW}), however, notably with the exception of  the fourth term \cite{foldy50}. The higher-order terms in the original FW transformation are of doubtful value \cite{Eriksen1958,DEVRIES1968,Silenko2013}. Therefore, the Hamiltonian in Eq.\ (\ref{originalHFW}) is not trustable and corrections are needed to achieve the expected higher-order terms. The main problem with the original FW transformation is that the unitary operators in two preceding transformations do not commute with each other. For example, for the exponential operators $e^{iU_{\rm FW}}$ and $e^{iU^\prime_{\rm FW}}$, the commutator $\left[U_{\rm FW},U_{\rm FW}^\prime\right]\neq 0$. Moreover, as the unitary operators are odd, this commutator produces even terms that have not been considered in the original FW transformation \cite{Silenko2016,Silenko2016PRA,Mondal2017thesis}.  Taking into account those terms, the correction of the FW transformation generates the Hamiltonian as \cite{Silenko2016}
\begin{align}
	\mathcal{H}_{\rm FW} ^{\rm corr.} & = (\underline{\beta} -\underline{\mathbbm{1}})mc^2 + \underline{\beta}\left(\frac{\mathcal{O}^2}{2mc^2}-\frac{\mathcal{O}^4}{8m^3c^6} + \frac{\mathcal{O}^6}{16m^5c^{10}}\right) + \mathcal{E} -  \frac{1}{8m^2c^4}\left[\mathcal{O},\left[\mathcal{O},\mathcal{F}\right]\right]  \nonumber\\
& + \frac{\underline{\beta}}{16m^3c^6}\left\{\mathcal{O},\left[\left[\mathcal{O},\mathcal{F}\right],\mathcal{F}\right]\right\} + \frac{3}{64m^4c^8}\left\{\mathcal{O}^2,\left[\mathcal{O},\left[\mathcal{O},\mathcal{F}\right]\right]\right\}+ \frac{1}{128m^4c^8}\left[\mathcal{O}^2,\left[\mathcal{O}^2,\mathcal{F}\right]\right]\nonumber\\
&  - \frac{1}{32m^4c^8}\left[\mathcal{O},\left[\left[\left[\mathcal{O},\mathcal{F}\right],\mathcal{F}\right],\mathcal{F}\right]\right] \,.
\label{corrected-FW} 
\end{align}
Note the difference between two Hamiltonians in Eq.\ (\ref{originalHFW}) and Eq.\ (\ref{corrected-FW}) that are observed in the second and consequent lines in both the equations, however, the terms in the first line are the same. Eq.\ (\ref{corrected-FW}) provides the correct higher-order terms of the FW transformation.
In this regard, we mention that an another approach towards the correct FW transformation has been employed by Eriksen; this is a single step approach that produces the expected FW transformed higher-order terms \cite{Eriksen1958}. 
Once the transformed Hamiltonian has been obtained as a function of odd and even terms, the final form is achieved by substituting the correct form of odd terms $\mathcal{O}$ and even terms $\mathcal{E}$ in the expression of Eq.\ (\ref{corrected-FW}) and calculating term by term. 

Since we perform here the time-dependent FW transformation, we note that the commutator $\left[\mathcal{O},\mathcal{F}\right] $ can be evaluated as $\left[\mathcal{O},\mathcal{F}\right] = i\hbar\, \partial\mathcal{O}/\partial t$. Therefore, following the definition of the odd operator, the time-varying fields are  taken into account through this term.
We evaluate each of the terms in Eq.\ (\ref{corrected-FW}) separately and obtain that the particles can be described by the following extended Pauli Hamiltonian \cite{hinschberger12,Mondal2017Nutation,Mondal2017thesis}
\begin{align}
\mathcal{H}^{\rm corr.}_{\rm FW} & =  \frac{\left(\bm{p}-e\bm{A}\right)^{2}}{2m}+V-\frac{e\hbar}{2m}\bm{\sigma}\cdot\bm{B}-\frac{\left(\bm{p}-e\bm{A}\right)^{4}}{8m^{3}c^{2}} +\frac{\left(\bm{p}-e\bm{A}\right)^{6}}{16m^{5}c^{4}}
 \nonumber\\
 &  - \left(\frac{e\hbar}{2m}\right)^2\frac{\bm{B}^2}{2mc^2} + \frac{e\hbar}{4m^2c^2}\left\{ \frac{\left(\bm{p}-e\bm{A}\right)^{2}}{2m}, \bm{\sigma}\cdot\bm{B}\right\} \nonumber\\
&-\frac{e\hbar^{2}}{8m^{2}c^{2}}\bm{\nabla}\cdot\bm{E}_{\rm tot} - \frac{e\hbar}{8m^{2}c^{2}}\bm{\sigma}\cdot\left[ \bm{E}_{\rm tot}\times\left(\bm{p}-e\bm{A}\right)-\left(\bm{p}-e\bm{A}\right)\times\bm{E}_{\rm tot}\right] \nonumber\\
&  - \frac{e\hbar^2}{16m^3c^4}\left\{\left(\bm{p}-e\bm{A}\right),\frac{\partial\bm{E}_{\rm tot}}{\partial t} \right\}- \frac{ie\hbar^2}{16m^3c^4}\bm{\sigma}\cdot\left[ \frac{\partial\bm{E}_{\rm tot}}{\partial t}\times\left(\bm{p}-e\bm{A}\right)+\left(\bm{p}-e\bm{A}\right)\times\frac{\partial\bm{E}_{\rm tot}}{\partial t}\right]\nonumber\\
& + \frac{3e\hbar}{64m^4c^4} \Big\{\left(\bm{p}-e\bm{A}\right)^2 -e\hbar\, \bm{\sigma}\cdot\bm{B},\hbar \bm{\nabla}\cdot\bm{E}_{\rm tot} +\bm{\sigma}\cdot\left[ \bm{E}_{\rm tot}\times\left(\bm{p}-e\bm{A}\right)-\left(\bm{p}-e\bm{A}\right)\times\bm{E}_{\rm tot}\right] \Big\} \nonumber\\
& + \frac{e\hbar^{4}}{32m^{4}c^{6}} \bm{\nabla}\cdot \frac{\partial^2\bm{E}_{\rm tot}}{\partial t^2} + \frac{e\hbar^{3}}{32m^{4}c^{6}}\bm{\sigma}\cdot\left[ \frac{\partial^2\bm{E}_{\rm tot}}{\partial t^2}\times\left(\bm{p}-e\bm{A}\right)-\left(\bm{p}-e\bm{A}\right)\times\frac{\partial^2\bm{E}_{\rm tot}}{\partial t^2}\right] .\,
\label{final_FW}
\end{align}
The fields in the last Hamiltonian (\ref{final_FW}) are defined as $\bm{B}=\bm{\nabla}\times\bm{A}$, the external magnetic field, $\bm{E}_{\rm tot} = \bm{E}_{\rm int} + \bm{E}_{\rm ext}$ are the electric fields where $\bm{E}_{\rm int}= -\frac{1}{e}\bm{\nabla}V$ is the internal field that exists even without any perturbation and $\bm{E}_{\rm ext}=-\frac{\partial\bm{A}}{\partial t}$ is the external field (only the temporal part is retained here because of the Coulomb gauge). 
It is clear that as the internal field is time-independent, it does not contribute to the fourth and sixth lines of Eq.\ (\ref{final_FW}). However, the external field does contribute to the above terms wherever it appears in the Hamiltonian.

The above-derived Hamiltonian can be split in two parts: (1) a spin-independent Hamiltonian and (2) a spin-dependent Hamiltonian that involves the Pauli spin matrices. The spin-dependent Hamiltonian, furthermore, has two types of coupling terms. The direct field-spin coupling terms are those which directly couples the fields with the magnetic moments e.g., the third term in the first line, the second term in the third line of Eq.\ (\ref{final_FW}) etc. 
On the other hand, there are relativistic terms that do not directly couple the spins to the electromagnetic field - indirect field-spin coupling terms. These terms include e.g., the second term of the second line, the fifth line of Eq.\ (\ref{final_FW}) etc.
The direct field-spin interaction terms are most important because these govern the directly manipulation of the spins in a system with an electromagnetic field. For the external electric field, these terms can be written together as a function of electric and magnetic field. These terms are taken into account and discussed in the next section.
The indirect coupling terms are often not taken into consideration and not included in the discussion (see Ref.\ \cite{hinschberger12,Zawadzki2005} for details).   
In this context, we reiterate that our current approach of deriving relativistic terms does not include the exchange and correlation effect. A similar FW transformed Hamiltonian has previously been derived, however, with a general Kohn-Sham exchange field \cite{Mondal2015a,Mondal2016,Mondal2017thesis}. As mentioned before, in this article we do not intend to include the exchange-correlation effect, while mostly focussing on the magnetic relaxation and magnetic inertial dynamics.

\subsection{The spin Hamiltonian}

The aim of this work is to formulate the spin dynamics on the basis of the Hamiltonian in Eq.\ (\ref{final_FW}). 
The direct field-spin interaction terms can be written together as electric or magnetic contributions. These two contributions can be expressed as a series up to an order of $1/m^5$  \cite{hinschberger12}
\begin{align}
	\mathcal{H}^{\bm{S}}_{\rm magnetic} & = - \frac{e}{m}\,\bm{S}\cdot \left[\bm{B} + \frac{1}{2} \sum_{n=1,2,3,4} \left(\frac{1}{2i\omega_c}\right)^n \frac{\partial^n\bm{B}}{\partial t^n}\right] + O \left(\frac{1}{m^6}\right)\,,
	\label{spin-Ham-2}
\end{align}
\begin{align}
	\mathcal{H}^{\bm{S}}_{\rm electric} & =- \frac{e}{m}\,\bm{S}\cdot \left[\frac{1}{2mc^2} \sum_{n=0,2} \left(\frac{i}{2\omega_c}\right)^n \frac{\partial^n\bm{E}}{\partial t^n}\times \left(\bm{p}-e\bm{A}\right)\right] + O \left(\frac{1}{m^6}\right)\,,
	\label{spin-Ham-1}
\end{align}
where the Compton wavelength and pulsation have been expressed by the usual definitions $\lambda_c = h /mc$ and $\omega_c = 2\pi c/\lambda_c$ with Plank's constant $h$.   
We also have used the spin angular momentum operator as $\bm{S}=(\hbar/2)\,\bm{\sigma}$. Note that we have dropped the notion of total electric field because the the involved fields ($\bm{B}$, $\bm{E}$, $\bm{A}$) are external only, the internal fields  are considered as time-independent.  
The involved terms in the above two spin-dependent Hamiltonians can readily be explained. The first term in the magnetic contribution in Eq.\ (\ref{spin-Ham-2}) explains the Zeeman coupling of spins to the external magnetic field. The rest of the terms in both the Hamiltonians in Eqs.\ (\ref{spin-Ham-1}) and (\ref{spin-Ham-2}) represent the spin-orbit coupling and its higher-order corrections. 
We note that these two spin Hamiltonians are individually not Hermitian, however, it can be shown that together they form a Hermitian Hamiltonian \cite{Mondal2015b}. 
As these Hamiltonians describe a semirelativistic Dirac particle, it is possible to derive from them the spin dynamics of a single Dirac particle \cite{Mondal2017Nutation}.  
The effect of the indirect field-spin terms is not yet well understood, but they could become important too in magnetism \cite{Zawadzki2005,hinschberger12}, however, those terms are not of our interest here.   

The electric Hamiltonian can be written in terms of magnetic contributions with the choice of a gauge $\bm{A}=\bm{B}\times\bm{r}/2$. The justification of the gauge lies in the fact that the magnetic field inside the system being studied is uniform \cite{Mondal2017thesis}. The transverse electric field in the Hamiltonian  (\ref{spin-Ham-2}) can be written as
\begin{align}
	\bm{E} = \frac{1}{2}\left(\bm{r}\times\frac{\partial\bm{B}}{\partial t}\right)\,.
\end{align}
Replacing this expression in the electric spin Hamiltonian in Eq.\ (\ref{spin-Ham-1}), one  can obtain a generalised expression of the total spin-dependent Hamiltonian as
\begin{align}
	\mathcal{H}^{\bm{S}}(t) & = - \frac{e}{m}\,\bm{S}\cdot \Big[\bm{B} + \frac{1}{2} \sum_{n=1,2,...}^{\infty} \left(\frac{1}{2i\omega_c}\right)^n \frac{\partial^n\bm{B}}{\partial t^n}\nonumber\\
	& + \frac{1}{4mc^2} \sum_{n=0,2,...}^{\infty} \left(\frac{i}{2\omega_c}\right)^n \left(\bm{r}\times\frac{\partial^{n+1}\bm{B}}{\partial t^{n+1}}\right)\times \left(\bm{p}-e\bm{A}\right)\Big] \,.
	\label{infinite-Hamil}
\end{align}
 It is important to stress that the above spin-Hamiltonian is a generalisation of the two Hamiltonians in Eqs.\ (\ref{spin-Ham-2}) and (\ref{spin-Ham-1}). We have already evaluated the Hamiltonian forms for $n=1,2,3,4$ and assume that the higher-order terms will have the same form \cite{hinschberger12}. 
This Hamiltonian consists of the direct field-spin interaction terms that are linear and/or quadratic in the fields. In the following we consider {\it only} the linear interaction terms, that is we neglect the $e\bm{A}$ term in Eq.\ (\ref{infinite-Hamil}). Here, we mention that the quadratic terms could provide an explanation towards the previously unknown  origin of spin-photon coupling or optical spin-orbit torque and angular magneto-electric coupling \cite{Mondal2015b,Mondal2017,Paillard2016SPIE}.
The linear direct field-spin Hamiltonian can then be recast as
\begin{align}
	\mathcal{H}^{\bm{S}}(t) & = - \frac{e}{m}\,\bm{S}\cdot \Big[\bm{B} + \frac{1}{2} \sum_{n=1,2,...}^{\infty} \left(\frac{1}{2i\omega_c}\right)^n \frac{\partial^n\bm{B}}{\partial t^n}\nonumber\\
	& + \frac{1}{4mc^2} \sum_{n=0,2,...}^{\infty} \left(\frac{i}{2\omega_c}\right)^n \left\{\frac{\partial^{n+1}\bm{B}}{\partial t^{n+1}} \left(\bm{r}\cdot\bm{p}\right) - \bm{r}\left(\frac{\partial^{n+1}\bm{B}}{\partial t^{n+1}}\cdot \bm{p}\right)\right\}\Big] \,.
	\label{spin-Ham-final}
\end{align}      
This is final form of the Hamiltonian and we are interested to describe to evaluate its contribution to the spin dynamics.
 
\section{Spin dynamics}

Once we have the explicit form of the spin Hamiltonian in Eq.\ (\ref{spin-Ham-final}), we can proceed to derive the corresponding classical magnetisation dynamics. Following similar procedures of previous work \cite{Mondal2016,Mondal2017Nutation}, and introducing a magnetisation element $\bm{M}(\bm{r},t)$, the magnetisation dynamics can be calculated by the following equation of motion
\begin{align}
	\frac{\partial\bm{M}}{\partial t} & = \sum_j \frac{g\mu_{\rm B}}{\Omega} \frac{1}{i\hbar}\Big{\langle} \big[\bm{S}^j,\mathcal{H}^{\bm{S}}(t)\big] \Big{\rangle}\,,
\end{align}
where $\mu_{\rm B}$ is the Bohr magneton, $g$ is the Land\'e g-factor that takes a value $\approx 2$ for electron spins and $\Omega$ is a suitably chosen volume element. Having the spin Hamiltonian in Eq.\ (\ref{spin-Ham-final}), we evaluate the corresponding commutators. As the spin Hamiltonian involves the magnetic fields, one can classify the magnetisation dynamics into two situations: (a) the system is driven by a harmonic field, (b) the system is driven by a non-harmonic field. However, in the below we continue the derivation of magnetisation dynamics with the harmonic driven fields. 
The magnetisation dynamics driven by the non-harmonic fields has been discussed in the context of Gilbert damping and inertial dynamics where it was shown that an additional torque contribution (the field-derivative torque) is expected to play a crucial role \cite{Mondal2016,Mondal2017Nutation,Mondal2017thesis}. 

The magnetisation dynamics due to the very first term of the Hamiltonian in Eq.\ (\ref{spin-Ham-final}) is derived as \cite{Mondal2017Nutation}
\begin{align}
	\frac{\partial\bm{M}^{(1)}}{\partial t} & = - \gamma \bm{M}\times\bm{B}\,, 
\end{align}  
with the gyromagnetic ratio $\gamma = g\vert e\vert/2m $. Here the commutators between two spin operators have been evaluated using $\left[S_j,S_k\right] = i\hbar S_l\epsilon_{jkl}$, where $\epsilon_{jkl}$ is the Levi-Civita tensor. This dynamics actually produces the precession of magnetisation vector around an effective field. To get the usual form of Landau-Lifshitz precessional dynamics, one has to use a linear relationship of magnetisation and magnetic field as $\bm{B}= \mu_0(\bm{M}+\bm{H})$. With the latter relation, the precessional dynamics becomes $-\gamma_0\bm{M}\times\bm{H}$, where $\gamma_0=\gamma\mu_0$ defines the effective gyromagnetic ratio. We point out that the there are relativistic contributions to the precession dynamics as well, e.g., from the spin-orbit coupling due to the time-independent field $\bm{E}_{\rm int}$ \cite{Mondal2016}. Moreover, the contributions to the magnetisation precession due to exchange field appear here, but are not explicitly considered in this article as they are not in the focus of the current investigations (see Ref.\ \cite{Mondal2016} for details).

The rest of the terms in the spin Hamiltonian in Eq.\ (\ref{spin-Ham-final}) is of much importance because they involve the time-variation of the magnetic induction. As it has been shown in an earlier work \cite{Mondal2016} that for the external fields and specifically the terms with $n=1$ in the second terms and $n=0$ in the third terms of Eq.\ (\ref{spin-Ham-final}), these terms together are Hermitian. 
These terms contribute to the magnetisation dynamics as the Gilbert relaxation within the LLG equation of motion, 
\begin{align}
	\frac{\partial\bm{M}^{(2)}}{\partial t} & = \bm{M}\times\left(A\cdot \frac{\partial\bm{M}}{\partial t}\right)\,,
\end{align}    
where the Gilbert damping parameter $A$ has been derived to be a tensor that has mainly two contributions: electronic and magnetic. The damping parameter $A$ has the form \cite{Mondal2017Nutation,Mondal2016}
\begin{align}
	A_{ij} & = - \frac{e\mu_0}{8m^2c^2} \sum_{\ell,k} \big[ \langle r_ip_k +p_kr_i \rangle - \langle r_{\ell} p_{\ell} + p_{\ell} r_{\ell} \rangle \delta_{ik}\big]\times \left(\mathbbm{1} + \chi^{-1}\right)_{kj}\,,
	\label{Gilbert}
\end{align}
where $\mathbbm{1}$ is the 3$\times 3$ unit matrix and  $\chi$ is the magnetic susceptibility tensor that can be introduced only if the system is driven by a field which is single harmonic \cite{Mondal2017thesis}. 
Note that the electronic contributions to the Gilbert damping parameter are given by the expectation value $\langle r_ip_k\rangle$ and the magnetic contributions by the susceptibility. We also mention that the tensorial Gilbert damping tensor has been shown to contain a scalar, isotropic Heisenberg-like contribution, an anisotropic Ising-like tensorial contribution and a chiral Dzyaloshinskii-Moriya-like contribution \cite{Mondal2016}. 

In an another work, we took into account the terms with $n=2$ in the second term of Eq.\ (\ref{spin-Ham-final}) and it has been shown that those containing the second-order time variation of the magnetic induction result in the magnetic inertial dynamics. Note that these terms provide a contribution to the higher-order relativistic effects. The corresponding magnetisation dynamics can be written as \cite{Mondal2017Nutation}  
\begin{align}
	\frac{\partial\bm{M}^{(3)}}{\partial t} & = \bm{M}\times\left(C\cdot \frac{\partial\bm{M}}{\partial t} + D\cdot\frac{\partial^2\bm{M}}{\partial t^2} \right)\,,
	\label{dm-dt3}
\end{align}    
with a higher-order Gilbert damping tensor $C_{ij}$ and inertia parameter $D_{ij}$ that have the following expressions $C_{ij} = \frac{\gamma_0\hbar^2}{8m^2c^4}\,\frac{\partial}{\partial t}(\mathbbm{1}+\chi^{-1})_{ij}$ and $D_{ij} = \frac{\gamma_0\hbar^2}{8m^2c^4} (\mathbbm{1} + \chi^{-1})_{ij}$.
We note that Eq.\ (\ref{dm-dt3}) contains two fundamentally different dynamics -- the first term on the right-hand side has the exact form of Gilbert damping dynamics whereas the second term has the form of magnetic inertial dynamics \cite{Mondal2017Nutation}.

The main aim of this article is to formulate a general magnetisation dynamics equation and an extension of the traditional LLG equation to include higher-order relativistic effects. The calculated magnetisation dynamics due to the second and third terms of Eq.\ (\ref{spin-Ham-final}) can be expressed as 
 \begin{align}
 	\frac{\partial\bm{M}}{\partial t}& =  \frac{e}{m}\,\bm{M}\times \Big[\frac{1}{2} \sum_{n=0,1,...}^{\infty} \left(\frac{1}{2i\omega_c}\right)^{n+1} \frac{\partial^{n+1}\bm{B}}{\partial t^{n+1}}\nonumber\\
	& + \frac{1}{4mc^2} \sum_{n=0,2,...}^{\infty} \left(\frac{i}{2\omega_c}\right)^n \left\{\frac{\partial^{n+1}\bm{B}}{\partial t^{n+1}} \langle\bm{r}\cdot\bm{p}\rangle - \Big\langle\bm{r}\left(\frac{\partial^{n+1}\bm{B}}{\partial t^{n+1}}\cdot \bm{p}\right)\Big\rangle\right\}\Big] \,.
	\label{dyn-with-B}
 \end{align}
Note the difference in the summation of first terms from the Hamiltonian in Eq.\ (\ref{spin-Ham-final}). 
To obtain explicit expressions for the Gilbert damping dynamics, we employ a general linear relationship between magnetisation and magnetic induction, $\bm{B}=\mu_0(\bm{H}+\bm{M})$.  
The time-derivative of the magnetic induction can then be replaced by magnetisation and magnetic susceptibility. For the $n$-th order time-derivative of the magnetic induction we find
\begin{align}
	\frac{\partial^n\bm{B}}{\partial t^n} & = \mu_0\left(\frac{\partial^n\bm{H}}{\partial t^n}+\frac{\partial^n\bm{M}}{\partial t^n}\right)\,.
\end{align}
Note that this equation is valid for the case when the magnetisation is time-dependent.
Substituting this expression into the Eq.\ (\ref{dyn-with-B}), one can derive the general LLG equation and its extensions. 
Moreover, as we work out the derivation in the case of harmonic driving fields, the differential susceptibility can be introduced as $\chi = \partial\bm{M}/\partial\bm{H}$.
The first term ($n$-th derivative of the magnetic field) can consequently be written by the following Leibniz formula as
\begin{align}
	\frac{\partial^n\bm{H}}{\partial t^n} & = \sum_{k=0}^{n-1}\frac{(n-1)!}{k! (n-k-1)!}\,\frac{\partial^{n-k-1}(\chi^{-1})}{\partial t^{n-k-1}}\cdot \frac{\partial^k}{\partial t^k}\left(\frac{\partial\bm{M}}{\partial t}\right)\,,
\end{align}  
where the magnetic susceptibility $\chi^{-1}$ is a time-dependent tensorial quantity and harmonic.
Using this relation, the first term and second terms in Eq.\ (\ref{dyn-with-B}) assume the form
\begin{align}
	\frac{\partial\bm{M}}{\partial t}\Big\vert _{\rm first} & =  \frac{e\mu_0}{2m}\,\bm{M}\times  \sum_{n=0,1,...}^{\infty} \left(\frac{1}{2i\omega_c}\right)^{n+1} 
	  \sum_{k=0}^{n} \frac{n!}{k! (n-k)!}\,\frac{\partial^{n-k}(\mathbbm{1}+\chi^{-1})}{\partial t^{n-k}}\cdot \frac{\partial^k}{\partial t^k}\left(\frac{\partial\bm{M}}{\partial t}\right)\,,
\label{first-term-dyn}
\end{align}
\begin{align}
	& \frac{\partial\bm{M}}{\partial t}\Big\vert _{\rm second} =  \frac{e\mu_0}{4m^2c^2}\,\bm{M}\times \nonumber\\ 
	& \sum_{n=0,2,...}^{\infty} \left(\frac{1}{2i\omega_c}\right)^n 
	  \sum_{k=0}^{n} \frac{n!}{k! (n-k)!}\,\Big[\frac{\partial^{n-k}(\mathbbm{1}+\chi^{-1})}{\partial t^{n-k}}\cdot \frac{\partial^k}{\partial t^k}\left(\frac{\partial\bm{M}}{\partial t}\right)\langle\bm{r}\cdot\bm{p}\rangle\nonumber\\
	& \qquad \qquad \qquad \qquad \qquad \qquad -\Big \langle \bm{r} \left(\left\{\frac{\partial^{n-k}(\mathbbm{1}+\chi^{-1})}{\partial t^{n-k}}\cdot \frac{\partial^k}{\partial t^k}\left(\frac{\partial\bm{M}}{\partial t}\right)\right\} \cdot \bm{p}\right) \Big\rangle  \Big].
\label{second-term-dyn}
\end{align}
These two equations already provide a generalisation of the higher-order magnetisation dynamics including the Gilbert damping (i.e., the terms with $k=0$) and the  inertial dynamics (the terms with $k=1$) and so on.  
  
\section{Discussion}

\subsection{Gilbert damping parameter}

It is obvious that, as Gilbert damping dynamics involves the first-order time derivative of the magnetisation and a torque due to it,  $k$ must take the value $k=0$ in the equations (\ref{first-term-dyn}) and (\ref{second-term-dyn}). Therefore, the Gilbert damping dynamics can be achieved from the following equations:
\begin{align}
	\frac{\partial\bm{M}}{\partial t}\Big\vert _{\rm first} & =  \frac{e\mu_0}{2m}\,\bm{M}\times\sum_{n=0,1,...}^{\infty}  \left(\frac{1}{2i\omega_c}\right)^{n+1}  \,\frac{\partial^{n}(\mathbbm{1}+\chi^{-1})}{\partial t^{n}}\cdot \frac{\partial\bm{M}}{\partial t}\,,
\end{align}     
\begin{align}
	& \frac{\partial\bm{M}}{\partial t}\Big\vert _{\rm second} =  \frac{e\mu_0}{4m^2c^2}\,\bm{M}\times \sum_{n=0,2,...}^{\infty} \left(\frac{1}{2i\omega_c}\right)^n 
	   \,\Big[\left(\frac{\partial^{n}(\mathbbm{1}+\chi^{-1})}{\partial t^{n}}\cdot \frac{\partial\bm{M}}{\partial t}\right)\langle\bm{r}\cdot\bm{p}\rangle\nonumber\\
	& \qquad \qquad \qquad \qquad \qquad \qquad -\Big \langle \bm{r} \left(\left\{\frac{\partial^{n}(\mathbbm{1}+\chi^{-1})}{\partial t^{n}}\cdot \frac{\partial\bm{M}}{\partial t}\right\} \cdot \bm{p}\right) \Big\rangle  \Big].
\end{align}
Note that these equations can be written in the usual form of Gilbert damping as $\bm{M}\times\left(\mathcal{G}\cdot \frac{\partial\bm{M}}{\partial t}\right)$, where the Gilbert damping parameter $\mathcal{G}$ is notably a tensor \cite{gilbert04,Mondal2016}. The general expression for the tensor can be given by a series of higher-order relativistic terms as follows
\begin{align}
	\mathcal{G}_{ij} & = \frac{e\mu_0}{2m}\sum_{n=0,1,...}^{\infty}  \left(\frac{1}{2i\omega_c}\right)^{n+1}  \,\frac{\partial^{n}(\mathbbm{1}+\chi^{-1})_{ij}}{\partial t^{n}}\nonumber\\
	& + \frac{e\mu_0}{4m^2c^2} \sum_{n=0,2,...}^{\infty} \left(\frac{1}{2i\omega_c}\right)^n 
	   \,\Big[\frac{\partial^{n}(\mathbbm{1}+\chi^{-1})_{ij}}{\partial t^{n}}\left(\langle r_l p_l\rangle -\langle r_lp_i\rangle \right)\Big]\,.
	   \label{Gilbert-series}
\end{align} 
Here we have used the Einstein summation convention on the index $l$. Note that there are two series: the first series runs over even and odd numbers ($n=0,\,1,\,2,\,3, \cdots$), however, the second series runs only over the even numbers ($n=0,\,2,\,4,Ê\cdots $). Eq.\ (\ref{Gilbert-series}) represents a general relativistic expression for the Gilbert damping tensor, given as a series of higher-order terms. This equation is one of the central results of this article.
It is important to observe that this expression provides the correct Gilbert tensor at the lowest relativistic order, i.e., putting $n=0$ the expression for the tensor is found to be exactly the same as Eq.\ (\ref{Gilbert}).

The analytic summation of the above series of higher-order relativistic contributions can be carried out when the susceptibility depends on the frequency of the harmonic driving field. 
This is in general true for ferromagnets where a differential susceptibility is introduced because there exists a spontaneous magnetisation in ferromagnets even without application of a harmonic external field. However, if the system is driven by a nonharmonic field, the introduction of the susceptibility is not valid anymore. In general the magnetic susceptibility is a function of wave vector and frequency in reciprocal space, i.e., $\chi = \chi(\bm{q},\omega)$. Therefore, for the single harmonic applied field, we use $\chi^{-1}\propto e^{i\omega t}$ 
 and the $n$-th order derivative will follow $\partial ^n/\partial t^n (\chi^{-1})\propto (i\omega)^n\chi^{-1}$. With these arguments, one can express the damping parameter of Eq.\ (\ref{Gilbert-series}) as (see Appendix A for detailed calculations) 
\begin{align}
	\mathcal{G}_{ij} 
	& =\frac{e\mu_0}{4m^2c^2}\left[\frac{\hbar}{i}+\langle r_l p_l\rangle -\langle r_lp_i\rangle\right] (\mathbbm{1}+\chi^{-1})_{ij}  \nonumber\\
	& + \frac{e\mu_0}{4m^2c^2}\left[\frac{(2\omega\omega_c + \omega^2)\frac{\hbar}{i} +\omega^2\left(\langle r_l p_l\rangle -\langle r_lp_i\rangle \right) }{4\omega_c^2-\omega^2}\right] \chi^{-1}_{ij}\,.
	\label{Gilbert-sum}
\end{align}  
Here, the first term in the last expression is exactly the same as the one that has been derived in our earlier investigation \cite{Mondal2016}. As the expression of  the expectation value $\langle r_ip_j\rangle$ is imaginary, the real Gilbert damping parameter will be given by the imaginary part of the susceptibility tensor. This holds consistently for the higher-order terms as well. The second term in Eq.\ (\ref{Gilbert-sum}) stems essentially from an infinite series which contain higher-order relativistic contributions to the Gilbert damping parameter. As $\omega_c$ scales with $c$, these higher-order terms will scale with $c^{-4}$ or more and thus their contributions will be smaller than the first term.
Note that the higher-order terms will diverge when $\omega=2\omega_c\approx 10^{21}$ sec$^{-1}$, which means that the theory breaks down at the limit $\omega \rightarrow 2\omega_c$. In this limit, the original FW transformation is not defined any more because the particles and antiparticles cannot be separated at this energy limit.  

\subsection{Magnetic inertia parameter}
Magnetic inertial dynamics, in contrast, involves a torque due to the second-order time-derivative of the magnetisation. In this case, $k$ must adopt the value $k=1$ in the afore-derived two equations (\ref{first-term-dyn}) and (\ref{second-term-dyn}). However, if $k=1$, the constraint $n-k\geq 0$ dictates that $n\geq 1$. Therefore, the magnetic inertial dynamics can be described with the following equations:
\begin{align}
	\frac{\partial\bm{M}}{\partial t}\Big\vert _{\rm first} & =  \frac{e\mu_0}{2m}\,\bm{M}\times  \sum_{n=1,2,...}^{\infty} \left(\frac{1}{2i\omega_c}\right)^{n+1} 
	   \frac{n!}{(n-1)!}\,\frac{\partial^{n-1}(\mathbbm{1}+\chi^{-1})}{\partial t^{n-1}}\cdot \frac{\partial^2\bm{M}}{\partial t^2}\,,
\end{align}
\begin{align}
	& \frac{\partial\bm{M}}{\partial t}\Big\vert _{\rm second} =  \frac{e\mu_0}{4m^2c^2}\,\bm{M}\times \sum_{n=2,4,...}^{\infty} \left(\frac{1}{2i\omega_c}\right)^n 
	   \frac{n!}{(n-1)!}\,\Big[\left(\frac{\partial^{n-1}(\mathbbm{1}+\chi^{-1})}{\partial t^{n-1}}\cdot \frac{\partial^2\bm{M}}{\partial t^2}\right)\langle\bm{r}\cdot\bm{p}\rangle\nonumber\\
	& \qquad \qquad \qquad \qquad \qquad \qquad -\Big \langle \bm{r} \left(\left\{\frac{\partial^{n-k}(\mathbbm{1}+\chi^{-1})}{\partial t^{n-k}}\cdot \frac{\partial^2\bm{M}}{\partial t^2}\right\} \cdot \bm{p}\right) \Big\rangle  \Big].
\end{align}
Similar to the Gilbert damping dynamics, these dynamical terms can be expressed as $\bm{M}\times \left(\mathcal{I}\cdot\frac{\partial^2\bm{M}}{\partial t^2} \right)$ which is the magnetic inertial dynamics \cite{Ciornei2011}. The corresponding parameter has the following expression
\begin{align}
	\mathcal{I}_{ij} & = \frac{e\mu_0}{2m}\sum_{n=1,2,...}^{\infty} \left(\frac{1}{2i\omega_c}\right)^{n+1} 
	   \frac{n!}{(n-1)!}\,\frac{\partial^{n-1}(\mathbbm{1}+\chi^{-1})_{ij}}{\partial t^{n-1}}\nonumber\\
	   & + \frac{e\mu_0}{4m^2c^2} \sum_{n=2,4,...}^{\infty} \left(\frac{1}{2i\omega_c}\right)^n 
	   \frac{n!}{(n-1)!}\,\Big[\frac{\partial^{n-1}(\mathbbm{1}+\chi^{-1})_{ij}}{\partial t^{n-1}}\left(\langle r_lp_l\rangle - \langle r_ip_l \rangle \right)\Big]\,.
	   \label{inertial-parameter-general}
\end{align}  
Note that as $n$ cannot adopt the value $n=0$, the starting values of $n$ are different in the two terms. Importantly, if $n=1$ we recover the expression for the lowest order magnetic inertia parameter $D_{ij}$, as given in the equation (\ref{dm-dt3}) \cite{Mondal2017Nutation}.  

Using similar arguments as in the case of the generalised Gilbert damping parameter, when we consider a single harmonic field as driving field, the inertia parameter can be rewritten as follows (see Appendix A for detailed calculations)
\begin{align}
	\mathcal{I}_{ij} 
	   & = -\frac{e\mu_0\hbar^2}{8m^3c^4} (\mathbbm{1}+\chi^{-1})_{ij}-\frac{e\mu_0\hbar^2}{8m^3c^4} \left(\frac{-\omega^2+4\omega\omega_c}{(2\omega_c-\omega)^2}\right)\chi^{-1}_{ij}\nonumber\\
	   & + \frac{e\mu_0}{8m^3c^4}\frac{\hbar}{i}\left(\langle r_lp_l\rangle - \langle r_ip_l \rangle \right)\left(\frac{16\omega\omega_c^3}{(4\omega_c^2-\omega^2)^2}\right)\chi^{-1}_{ij}\,.
	   \label{inertia-sum}
\end{align}  
The first term here is exactly the same as the one that was obtained in our earlier investigation \cite{Mondal2017Nutation}. However, there are now two extra terms which depend on the frequency of the driving field and that vanish for $\omega \rightarrow 0$. Again, in the limit $\omega\rightarrow2\omega_c$, these two terms diverge and hence this expression is not valid anymore. 
The inertia parameter will consistently be given by the real part of the susceptibility.

\section{Summary}
We have developed a generalised LLG equation of motion starting from fundamental quantum relativistic theory. Our approach leads to higher-order relativistic correction terms in the equation of spin dynamics of Landau and Lifshitz. To achieve this, we have started from the foundational Dirac equation under the presence of an electromagnetic field (e.g., external driving fields or THz excitations) and have employed the FW transformation to separate out the particles from the antiparticles in the Dirac equation. In this way, we derive an extended Pauli Hamiltonian which efficiently describes the interactions between the quantum spin-half particles and the applied field. The thus-derived direct field-spin interaction Hamiltonian can be generalised for any higher-order relativistic corrections and has been expressed as a series. To derive the dynamical equation, we have used this generalised spin Hamiltonian to calculate the corresponding spin dynamics using the Heisenberg equation of motion.
The obtained spin dynamical equation provides a generalisation of the phenomenological LLG equation of motion and moreover, puts the LLG equation on a rigorous foundational footing. The equation includes all the torque terms of higher-order time-derivatives of the magnetisation (apart from the Gilbert damping and magnetic inertial dynamics).
Specifically, however, we have focussed on deriving an analytic expression for the generalised Gilbert damping and for the magnetic inertial parameter. Our results show that both these parameters can be expressed as a series of higher-order relativistic contributions and that they are tensors. 
These series can be summed up for the case of a harmonic driving field, leading to closed analytic expressions. We have further  shown that the imaginary part of the susceptibility contributes to the Gilbert damping parameter 
 while the real part contributes to the magnetic inertia parameter. Lastly, with respect to the applicability limits of the derived expressions we have pointed out that when the frequency of the driving field becomes comparable to the Compton pulsation, our theory will not be valid anymore because of the spontaneous particle-antiparticle pair-production.

\section{Acknowledgments}

We thank P-A.\ Hervieux for valuable discussions.
This work has been supported by the Swedish Research Council (VR), the Knut and Alice Wallenberg Foundation (Contract No.\ 2015.0060), the European Union's Horizon2020 Research and Innovation Programme under grant agreement No.\ 737709 (FEMTOTERABYTE, {\blue http://www.physics.gu.se/femtoterabyte}).

\newpage  

\appendix
\section{Detailed calculations of the parameters for a harmonic field}
In the following we provide the calculational details of the summation towards the results given in Eqs.\ (\ref{Gilbert-sum}) and (\ref{inertia-sum}).
\subsection{Gilbert damping parameter}
Eq.\ (\ref{Gilbert-series}) can be expanded as follows
\begin{align}
	\mathcal{G}_{ij} & = \frac{e\mu_0}{2m}  \frac{1}{2i\omega_c}(\mathbbm{1}+\chi^{-1})_{ij} + \frac{e\mu_0}{4m^2c^2}  
	  \left(\langle r_l p_l\rangle -\langle r_lp_i\rangle \right)(\mathbbm{1}+\chi^{-1})_{ij} \nonumber\\
	& + \frac{e\mu_0}{2m}\sum_{n=1,2,...}^{\infty}  \left(\frac{1}{2i\omega_c}\right)^{n+1}  (i\omega)^n\chi^{-1}_{ij} + \frac{e\mu_0}{4m^2c^2} \sum_{n=2,4,...}^{\infty} \left(\frac{1}{2i\omega_c}\right)^n 
	   \left(\langle r_l p_l\rangle -\langle r_lp_i\rangle \right)(i\omega)^n\chi^{-1}_{ij}\nonumber\\
	   & =\frac{e\mu_0}{2m}  \frac{1}{2i\omega_c}(\mathbbm{1}+\chi^{-1})_{ij} + \frac{e\mu_0}{4m^2c^2}  
	  \left(\langle r_l p_l\rangle -\langle r_lp_i\rangle \right)(\mathbbm{1}+\chi^{-1})_{ij} \nonumber\\
	& + \frac{e\mu_0}{2m}\frac{1}{2i\omega_c}\sum_{n=1,2,...}^{\infty}  \left(\frac{\omega}{2\omega_c}\right)^{n}  \chi^{-1}_{ij} + \frac{e\mu_0}{4m^2c^2} \sum_{n=2,4,...}^{\infty} \left(\frac{\omega}{2\omega_c}\right)^n 
	   \left(\langle r_l p_l\rangle -\langle r_lp_i\rangle \right)\chi^{-1}_{ij}\nonumber\\
	   & =\frac{e\mu_0}{4m^2c^2}\left[\frac{\hbar}{i}+\langle r_l p_l\rangle -\langle r_lp_i\rangle\right] (\mathbbm{1}+\chi^{-1})_{ij}  \nonumber\\
	& + \frac{e\mu_0}{4m^2c^2}\left[\frac{\hbar}{i}\sum_{n=1,2,...}^{\infty}  \left(\frac{\omega}{2\omega_c}\right)^{n} + \left(\langle r_l p_l\rangle -\langle r_lp_i\rangle \right) \sum_{n=2,4,...}^{\infty} \left(\frac{\omega}{2\omega_c}\right)^n 
	   \right] \chi^{-1}_{ij}\nonumber\\
	   & =\frac{e\mu_0}{4m^2c^2}\left[\frac{\hbar}{i}+\langle r_l p_l\rangle -\langle r_lp_i\rangle\right] (\mathbbm{1}+\chi^{-1})_{ij}  \nonumber\\
	& + \frac{e\mu_0}{4m^2c^2}\left[\frac{\hbar}{i} \frac{\omega}{2\omega_c - \omega}+ \left(\langle r_l p_l\rangle -\langle r_lp_i\rangle \right) \frac{\omega^2}{4\omega_c^2-\omega^2}\right] \chi^{-1}_{ij}\nonumber\\
	& =\frac{e\mu_0}{4m^2c^2}\left[\frac{\hbar}{i}+\langle r_l p_l\rangle -\langle r_lp_i\rangle\right] (\mathbbm{1}+\chi^{-1})_{ij}  \nonumber\\
	& + \frac{e\mu_0}{4m^2c^2}\left[\frac{(2\omega\omega_c + \omega^2)\frac{\hbar}{i} +\omega^2\left(\langle r_l p_l\rangle -\langle r_lp_i\rangle \right) }{4\omega_c^2-\omega^2}\right] \chi^{-1}_{ij}\,.
\end{align}  
We have used the fact that $\frac{\omega}{\omega_c}<1$ and the summation formula 
\begin{align}
	1+x+x^2+x^3+... = \frac{1}{1-x}; \qquad -1<x<1\,.
\end{align}

\subsection{Magnetic inertia parameter}
Eq.\ (\ref{inertial-parameter-general}) can be expanded as follows
\begin{align}
	\mathcal{I}_{ij} & = \frac{e\mu_0}{2m}\left(\frac{1}{2i\omega_c}\right)^{2} (\mathbbm{1}+\chi^{-1})_{ij}+\frac{e\mu_0}{2m}\sum_{n=2,3,...}^{\infty} \left(\frac{1}{2i\omega_c}\right)^{n+1} 
	   \frac{n!}{(n-1)!}\,\frac{\partial^{n-1}(\mathbbm{1}+\chi^{-1})_{ij}}{\partial t^{n-1}}\nonumber\\
	   & + \frac{e\mu_0}{4m^2c^2} \sum_{n=2,4,...}^{\infty} \left(\frac{1}{2i\omega_c}\right)^n 
	   \frac{n!}{(n-1)!}\,\Big[\frac{\partial^{n-1}(\mathbbm{1}+\chi^{-1})_{ij}}{\partial t^{n-1}}\left(\langle r_lp_l\rangle - \langle r_ip_l \rangle \right)\Big]\nonumber\\
	   & = \frac{e\mu_0}{2m}\left(\frac{1}{2i\omega_c}\right)^{2} (\mathbbm{1}+\chi^{-1})_{ij}+\sum_{n=2,3,...}^{\infty} \left(\frac{1}{2i\omega_c}\right)^{n+1} 
	   \frac{n!}{(n-1)!}\,(i\omega)^{n-1}\chi^{-1}_{ij}\nonumber\\
	   & + \frac{e\mu_0}{4m^2c^2} \sum_{n=2,4,...}^{\infty} \left(\frac{1}{2i\omega_c}\right)^n 
	   \frac{n!}{(n-1)!}\,\left(\langle r_lp_l\rangle - \langle r_ip_l \rangle \right)(i\omega)^{n-1}\chi^{-1}_{ij}\nonumber\\
	   & = -\frac{e\mu_0\hbar^2}{8m^3c^4} (\mathbbm{1}+\chi^{-1})_{ij}+\frac{e\mu_0}{2m}\left(\frac{1}{2i\omega_c}\right)^2\sum_{n=2,3,...}^{\infty} \left(\frac{\omega}{2\omega_c}\right)^{n-1} 
	   \frac{n!}{(n-1)!}\,\chi^{-1}_{ij}\nonumber\\
	   & + \frac{e\mu_0}{4m^2c^2} \left(\frac{1}{2i\omega_c}\right)\sum_{n=2,4,...}^{\infty} \left(\frac{\omega}{2\omega_c}\right)^{n-1} 
	   \frac{n!}{(n-1)!}\,\left(\langle r_lp_l\rangle - \langle r_ip_l \rangle \right)\chi^{-1}_{ij}\nonumber\\
	   & = -\frac{e\mu_0\hbar^2}{8m^3c^4} (\mathbbm{1}+\chi^{-1})_{ij}-\frac{e\mu_0\hbar^2}{8m^3c^4}\sum_{n=2,3,...}^{\infty} \left(\frac{\omega}{2\omega_c}\right)^{n-1} 
	   \frac{n!}{(n-1)!}\,\chi^{-1}_{ij}\nonumber\\
	   & + \frac{e\mu_0}{8m^3c^4}\frac{\hbar}{i}\sum_{n=2,4,...}^{\infty} \left(\frac{\omega}{2\omega_c}\right)^{n-1} 
	   \frac{n!}{(n-1)!}\,\left(\langle r_lp_l\rangle - \langle r_ip_l \rangle \right)\chi^{-1}_{ij}\nonumber\\
	   & = -\frac{e\mu_0\hbar^2}{8m^3c^4} (\mathbbm{1}+\chi^{-1})_{ij}-\frac{e\mu_0\hbar^2}{8m^3c^4}\chi^{-1}_{ij} \left[2\left(\frac{\omega}{2\omega_c}\right)+3\left(\frac{\omega}{2\omega_c}\right)^2+4\left(\frac{\omega}{2\omega_c}\right)^3+...\right]\nonumber\\
	   & + \frac{e\mu_0}{8m^3c^4}\frac{\hbar}{i}\left(\langle r_lp_l\rangle - \langle r_ip_l \rangle \right)\chi^{-1}_{ij}\left[2\left(\frac{\omega}{2\omega_c}\right)+4\left(\frac{\omega}{2\omega_c}\right)^3+ 6\left(\frac{\omega}{2\omega_c}\right)^5+...\right]\nonumber\\
	   & = -\frac{e\mu_0\hbar^2}{8m^3c^4} (\mathbbm{1}+\chi^{-1})_{ij}-\frac{e\mu_0\hbar^2}{8m^3c^4} \left(\frac{-\omega^2+4\omega\omega_c}{(2\omega_c-\omega)^2}\right)\chi^{-1}_{ij}\nonumber\\
	   & + \frac{e\mu_0}{8m^3c^4}\frac{\hbar}{i}\left(\langle r_lp_l\rangle - \langle r_ip_l \rangle \right)\left(\frac{16\omega\omega_c^3}{(4\omega_c^2-\omega^2)^2}\right)\chi^{-1}_{ij}\,.
\end{align}  
Here we have used the formula
\begin{align}
	1+2x+3x^2+4x^3+5x^4+... = \frac{1}{(1-x)^2}; \qquad -1<x<1 \,.
\end{align}

\providecommand{\newblock}{}

\end{document}